\documentclass{SCGE}

\begin{document}

\begin{picture}(0,0){\rm
\put(0,-39){\makebox[160truemm][l]{\bf {\sanhao\raisebox{2pt}{.}}
Article  {\sanhao\raisebox{1.5pt}{.}}}}}
\put(0,-52){\jiuwuhao {\textcolor[rgb]{0.5,0.5,0.5}{\sf 
}}}
\end{picture}

\def\bm{\boldsymbol}

\def\dl{\displaystyle}
\def\du{\end{document}}
\def\pi{{\uppi}}

\Year{0000} %
\Month{00} %
\Vol{00} %
\No{0} %
\BeginPage{000} %
\EndPage{000} %
\AuthorMark{{\rm Chen Y.}}  %
\AuthorMarkCite{{\rm Chen Y.}.} %
\DOI{*} 

\title{Solar Ultraviolet Bursts in a Coordinated Observation of IRIS, Hinode and SDO}%

\author[1]{Yajie Chen}{}
\address[{1\rm}]{School of Earth and Space Sciences, Peking University, Beijing 100871, China}

\author[1*]{Hui Tian}{}
\footnote{*Hui Tian (email: huitian@pku.edu.cn)}

\author[2]{Xiaoshuai Zhu}{}
\address[{2\rm}]{Max Planck Institute for Solar System Research, Justus-von-Liebig-Weg 3, 37077, G\"{o}ttingen, Germany}

\author[1]{Tanmoy Samanta}{}

\author[1]{Linghua Wang}{}

\author[1]{Jiansen He}{}

\maketitle \vspace{-3.5mm}{\footnotesize\begin{center} Received ; accepted ; published online 
\end{center}}\vspace*{-5mm}

\begin{center}
\rule{16.5cm}{0.4pt}
\parbox{16.5cm}
{\begin{abstract} 

Solar ultraviolet (UV) bursts are small-scale compact brightenings in transition region images. The spectral profiles of transition region lines in these bursts are significantly enhanced and broadened, often with chromospheric absorption lines such as Ni~{\sc{ii}} 1335.203 and 1393.330 {\AA} superimposed. We investigate the properties of several UV bursts using a coordinated observation of the Interface Region Imaging Spectrograph (IRIS), Solar Dynamics Observatory (SDO), and \textit{Hinode} on 2015 February 7. We have identified 12 UV bursts, and 11 of them reveal small blueshifts of the Ni~{\sc{ii}} absorption lines. However, the Ni~{\sc{ii}} lines in one UV burst exhibit obvious redshifts of $\sim$20 km s$^{-1}$, which appear to be related to the cold plasma downflows observed in the IRIS slit-jaw images. We also examine the three-dimensional magnetic field topology using a magnetohydrostatic model, and find that some UV bursts are associated with magnetic null points or bald patches. In addition, we find that these UV bursts reveal no obvious coronal signatures from the observations of the Atmospheric Imaging Assembly (AIA) on board SDO and the EUV Imaging Spectrometer (EIS) on board \textit{Hinode}.

\end{abstract}}
\end{center}\vspace*{-0.6cm}

\begin{center}
\parbox{16.5cm}
{\bf\jiuhao Sun: chromosphere, Sun: transition region, Sun: UV radiation, magnetic reconnection}%
\end{center}

\begin{center}
{\PACS{\rm *}}
\Cit{*}
\end{center}

\wuhao\vspace*{1.5mm}

\begin{multicols}{2}

\renewcommand{\baselinestretch}{1.08} \baselineskip 12.2pt\parindent=10.8pt

\renewcommand{\thefootnote}

\section{Introduction}\vspace*{-2mm}

Recent observations from the Interface Region Imaging Spectrometer (IRIS) [1] have revealed a new type of magnetic reconnection events in the lower solar atmosphere called ``UV bursts" [2,3]. Some authors named them ``IRIS bombs" in earlier studies [4,5]. These events exhibit very intense and compact brightenings in the solar transition region images taken by IRIS. They are characterized by significantly enhanced and broadened profiles of some Si~{\sc{iv}} and C~{\sc{ii}} lines, with the superposition of some chromospheric absorption lines from the Ni~{\sc{ii}} and Fe~{\sc{ii}} ions on the profiles [2]. 

UV bursts are generally observed in emerging active regions [6,7,8,9,10]. They are usually associated with magnetic flux cancellations in the photosphere [2,7,10]. Zhao et al. [7] and Tian et al. [10] reconstructed three-dimensional magnetic field structures through magnetohydrostatic models based on the vector photospheric magnetograms obtained from the Helioseismic and Magnetic Imager (HMI) [11] on board the Solar Dynamics Observatory (SDO) [12], and found that some UV bursts are associated with bald patches [48]. Based on a nonlinear force-free-field (NLFFF) extrapolation, Chitta et al. [13] found an UV burst associated with the fan-spine magnetic field topology. These magnetic field extrapolations suggest that the heights of reconnection are generally in the range of 0.5--1.0 Mm, indicating that UV bursts are caused by magnetic reconnection in the lower solar atmosphere.

There is another type of reconnection events in the lower solar atmosphere called Ellerman bombs (EBs), which are characterized by compact transient brightenings in the H$\alpha$ wing images and no obvious signals in the H$\alpha$ core images [14,15,16,17,18,19,20,21,22,23]. Coordinated IRIS and ground-based observations revealed that some UV bursts are related to EBs [4,21,24]. The connection between UV bursts and EBs suggests that cool materials near the temperature minimum region (TMR) might be heated up to $\sim$80 kK (the formation temperature of the Si~{\sc{iv}} lines assuming ionization equilibrium) during the reconnection. The formation temperature of the Si~{\sc{iv}} lines could be 10--20 kK by assuming local thermodynamic equilibrium for line extinctions during the onsets of EBs [25]. However, existing one-dimensional models of EBs usually predict that the TMR could only be heated to less than 10 kK [5,26,27,28,29,30,31,32,33], which appears to be inconsistent with recent observations. On the other hand, the 2.5-dimensional single-fluid high resolution magnetohydrodynamic (MHD) simulations [34] and the further multi-fluid MHD results by including the non-equilibrium ionization-recombination effects [35,62] revealed that the materials around the TMR can be heated to a few tens of thousand Kelvin during the magnetic reconnection process if the plasma beta is low and the magnetic field is strong. Moreover, Hansteen et al. [36] reproduced UV bursts and EBs successfully in three-dimensional single fluid radiative MHD simulations with artificial hyper-diffusion. However, these two types of events take place at different locations and different times in the models. So their simulations can not explain the observed connection between UV bursts and EBs. 

Previous studies showed that most UV bursts reveal no obvious signatures in the coronal channels of the Atmospheric Imaging Assembly (AIA) [37] on board SDO. The absence of coronal emission in AIA coronal channels suggests that the plasmas in UV bursts are likely not heated to $\sim$$10^{6}$ K. However, recently Li et al. [38] identified signatures of a UV burst in the AIA coronal images, and concluded that the burst is heated to coronal temperatures. We have to mention that the AIA coronal passbands also have contribution from some strong transition region lines, and that a solid conclusion can only be made after examining pure coronal emission lines. 

In this paper, we present analysis results from a coordinated observation of IRIS, SDO, and \textit{Hinode} on 2015 February 7. We first identify all the UV bursts from the IRIS observation and examine the line profiles of the bursts. Then we investigate the magnetic field topology around the UV bursts using a magnetohydrostatic model. We also search for coronal signatures of the UV bursts using the observation from the EUV Imaging Spectrometer (EIS) [39] on board \textit{Hinode}.

\section{Observations}\vspace*{-2mm}

IRIS performed a very large dense raster (174$^{\prime\prime}$ along the slit, 400 raster steps with a step size of $\sim$0.$^{\prime\prime}$35) of NOAA active region (AR) 12280 from 04:10 to 04:46 UT. The pointing coordinate was (14$^{\prime\prime}$, 18$^{\prime\prime}$), close to the disk center. The cadence of the spectral observation was $\sim$5.4 s. The spatial pixel size of spectral images was  $\sim$0.$^{\prime\prime}$167 along the slit. The spectral dispersion was $\sim$0.013 per pixel in the far ultraviolet (FUV) and $\sim$0.026 {\AA} per pixel in the near ultraviolet (NUV) wavelength bands. The spatial pixel size was $\sim$0.$^{\prime\prime}$167 for slit-jaw images (SJI). Slit-jaw images were taken with the 1330, 1400, 2796, and 2832 {\AA} filters, and the cadences are 23 s, 23 s, 23 s, and 135 s, respectively. The exposure time of both the spectral and imaging observations was $\sim$4 s. We remove the solar rotation effect to internally coalign the SJI images,  and then use the fiducial line to coalign the images taken in different SJI filters and spectral windows.

The absolute wavelength calibration has been performed for the spectral data of IRIS following the methods described in Tian et al. [4]. We assume a zero average Doppler shift of the chromospheric Fe~{\sc{ii}} 1392.817 {\AA} line to calibrate the Si~{\sc{iv}} 1393.755 {\AA} spectral window. We also assume the same average Doppler velocity for the two Si~{\sc{iv}} lines to calibrate the Si~{\sc{iv}} 1402.770 {\AA} spectral window. For the C~{\sc{ii}} spectral window, we force the Ni~{\sc{ii}} 1335.203 and 1393.330 {\AA} lines to have the same average Doppler shift. The wavelength calibration for the Mg~{\sc{ii}} window is achieved by assuming that the average Doppler shifts of some strong neutral absorption lines are zero. After wavelength calibration, a single Gaussian fit is applied to the Si~{\sc{iv}} 1393.755 {\AA} line profiles to derive the intensity, Doppler velocity, and line width.

We also analyze the data taken by AIA and HMI onboard SDO. The cadences of AIA images were 12 s in the 131, 171, and 193 {\AA} passbands, and 24 s in the 1700 {\AA} passband. The pixel size of the AIA images was $\sim$0.$^{\prime\prime}$613. The AIA images in different passbands are coaligned by using the IDL routine \textit{aia\_prep.pro} available in \textit{SolarSoft}. In order to coalign AIA images and IRIS images, we compare the associated bright features in a Mg~{\sc{ii}} k wing image (sum of +1.33 {\AA} and $-$ 1.33 {\AA}, following the method in Tian et al. [4]) constructed from the IRIS spectral data with those in the first AIA 1700 {\AA} image. The HMI instrument provides vector magnetic field of the photosphere at a cadence of 720 s. The pixel size of HMI images was $\sim$0.$^{\prime\prime}$504. We use HMI vector magnetic field to study the magnetic field topologies of UV bursts.

We also use the data taken by the spectropolarimeter (SP) [40] of the \textit{Hinode}/Solar Optical Telescope (SOT) [41]. The SP performed a single raster scan from 03:46 to 04:41 UT in the Fe~{\sc{i}} 6301.5 and 6302.5 {\AA}~lines. The step size of the raster was $\sim$0.$^{\prime\prime}$30, and the pixel size along the slit was $\sim$0.$^{\prime\prime}$32. The SP level2 data (processed by full Milne-Eddington inversion from the level1 data) were directly downloaded from the instrument website. The coalignment between the SP data and the IRIS images is achieved by comparing the bright features in the SP continuum intensity image with the associated bright features in the Mg~{\sc{ii}} k wing image mentioned above. 

The EIS instrument performed sixty-two 20-step rasters from 02:03 to 05:41 UT, with a step size of $\sim$3$^{\prime\prime}$. Each raster lasted for $\sim$210 s. The exposure time was $\sim$9 s. The width of the slit was 2$^{\prime\prime}$. The spatial pixel size was $\sim$1$^{\prime\prime}$ along the slit, with a spectral dispersion of $\sim$0.0223 {\AA} per pixel. We first use the standard correction and calibration procedure to reduce the EIS data, and then perform a running average over three adjacent spatial pixels along the slit to increase the signal-to-noise ratio. The Fe~{\sc{x}} 184.54 {\AA}, Fe~{\sc{xii}} 195.12 {\AA}, Fe~{\sc{xiv}} 264.78 {\AA}, Fe~{\sc{xvi}} 262.98 {\AA}, and He~{\sc{ii}} 256.32 {\AA} lines are selected for our study. We apply a single Gaussian fit to each line profile to derive the line intensity. Then we coalign the EIS data and AIA images by comparing the AIA 193 {\AA} images and EIS Fe~{\sc{xii}} 195.12 {\AA} intensity images. This will also coalign the IRIS and EIS images, since the AIA and IRIS image have already been coaligned.

\section{Spectral characteristics of UV bursts}\vspace*{-2mm}

\begin{figure*} 
\centering {\includegraphics[width=\textwidth]{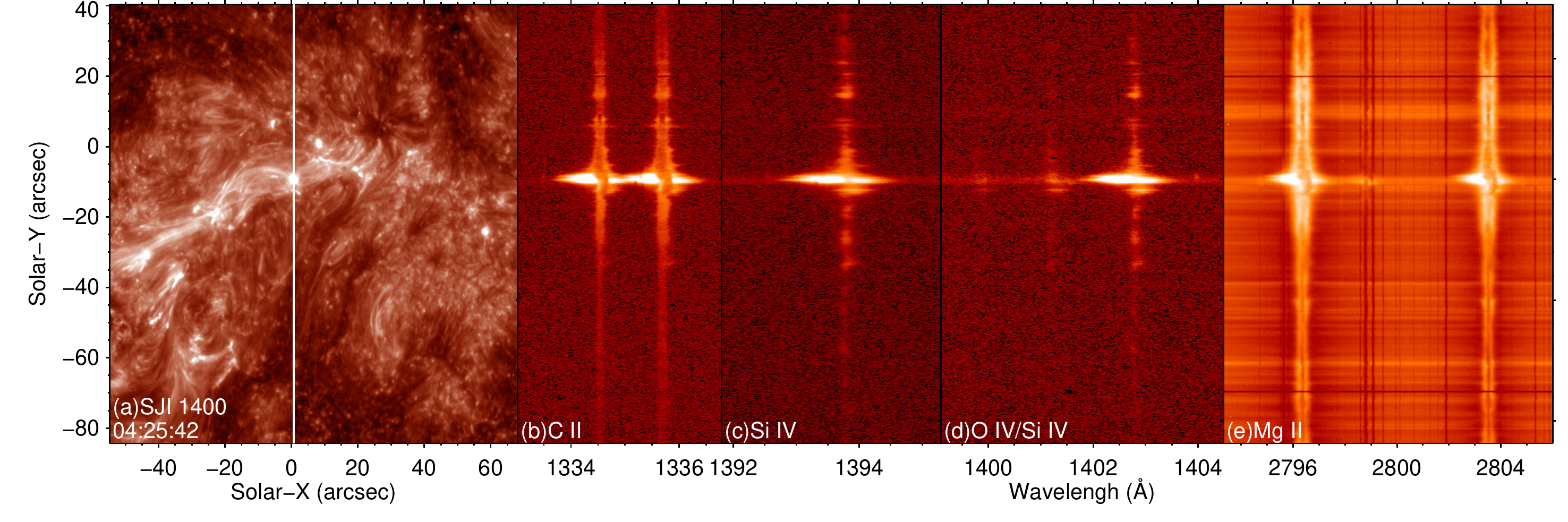}} 
\caption{A snapshot of IRIS imaging and spectral observation. Panel (a): SJI 1400 {\AA} image taken around 04:25:42 UT on 2015 Feb 7. The vertical white line marks the location of the slit. Panel (b)--(e): Simultaneous spectral images in four different spectral windows along the slit. The intensities are shown in arbitrary unit.
} \label{f1}
\end{figure*}

Figure ~\ref{f1} presents a 1400 {\AA} SJI image and spectral images in different spectral windows taken by IRIS around 04:25:42 UT. This figure only shows part of the whole field-of-view (FOV) of the IRIS observation. Some small-scale brightenings are visible in the SJI 1400 {\AA} image, and the slit crosses one of them at this time. The spectral profiles of the Si~{\sc{iv}}, C~{\sc{ii}}, and Mg~{\sc{ii}} lines are greatly broadened and enhanced at the location of the brightening. These broadened line profiles may be caused by magnetic reconnection [2,42,43]. In the C~{\sc{ii}} and Si~{\sc{iv}} 1393.755 {\AA} spectral windows, we can see some absorption signals of the Ni~{\sc{ii}} 1335.203 and 1393.330 {\AA} lines at the brightening, suggesting that this compact brightening is an UV burst. Then we try to identify all the UV bursts in this dataset. We first select all the small-scale brightenings in the intensity image of the Si~{\sc{iv}} 1393.755 {\AA} line. Then we examine the line profiles of each brightening. If the spectral profiles of the Si~{\sc{iv}} and C~{\sc{ii}} lines are greatly enhanced and broadened, and the Ni~{\sc{ii}} 1393.330 and 1335.203 {\AA} absorption lines are clearly present, we treat it as an UV burst. We have identified 12 UV bursts in this dataset. It is worth mentioning that the definition of UV burst is slightly different by different authors. For instance, UV bursts are defined mainly based on the significantly enhanced intensity by Young et al. [3]. So some events with good single Gaussian profiles and no absorption lines are also categorized into UV bursts (e.g., Hou et al. [66]). Here we use the definition in Tian et al. [4].

\begin{figure*} 
\centering {\includegraphics[width=\textwidth]{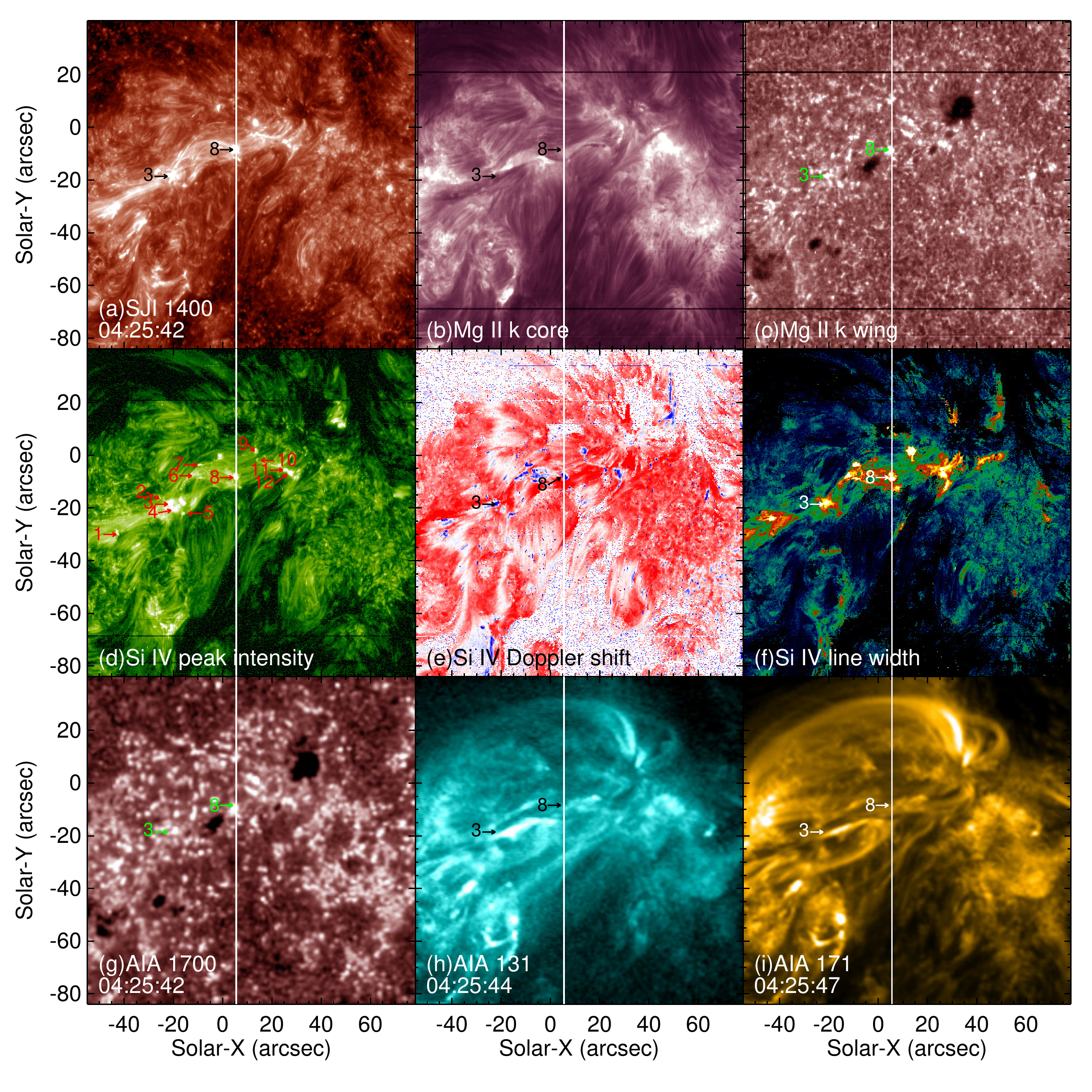}} 
\caption{Overview of the active region. (a): IRIS/SJI 1400 {\AA} image taken around 04:25:42 UT. (b) and (c): Mg~{\sc{ii}} k core and wing (sum of +/$-$ 1.33 {\AA}) images. (d)--(f): Images of the Si~{\sc{iv}} 1393.755 {\AA} intensity, Doppler velocity, and line width derived from a single Gaussian fit. The Doppler velocity saturates at $+/-$ 30 km~s$^{-1}$. (g)--(i) SDO/AIA 1700, 131, and 171 {\AA} images taken around 04:25:42, 04:25:44, 04:25:47 UT, respectively. The white vertical lines indicate the location of the slit. Identified UV bursts are marked in panel (d). The intensities are shown in arbitrary unit. The locations of bursts 3 and 8 are marked in all the panels.
} \label{f2}
\end{figure*}

\begin{figure*} 
\centering {\includegraphics[width=\textwidth]{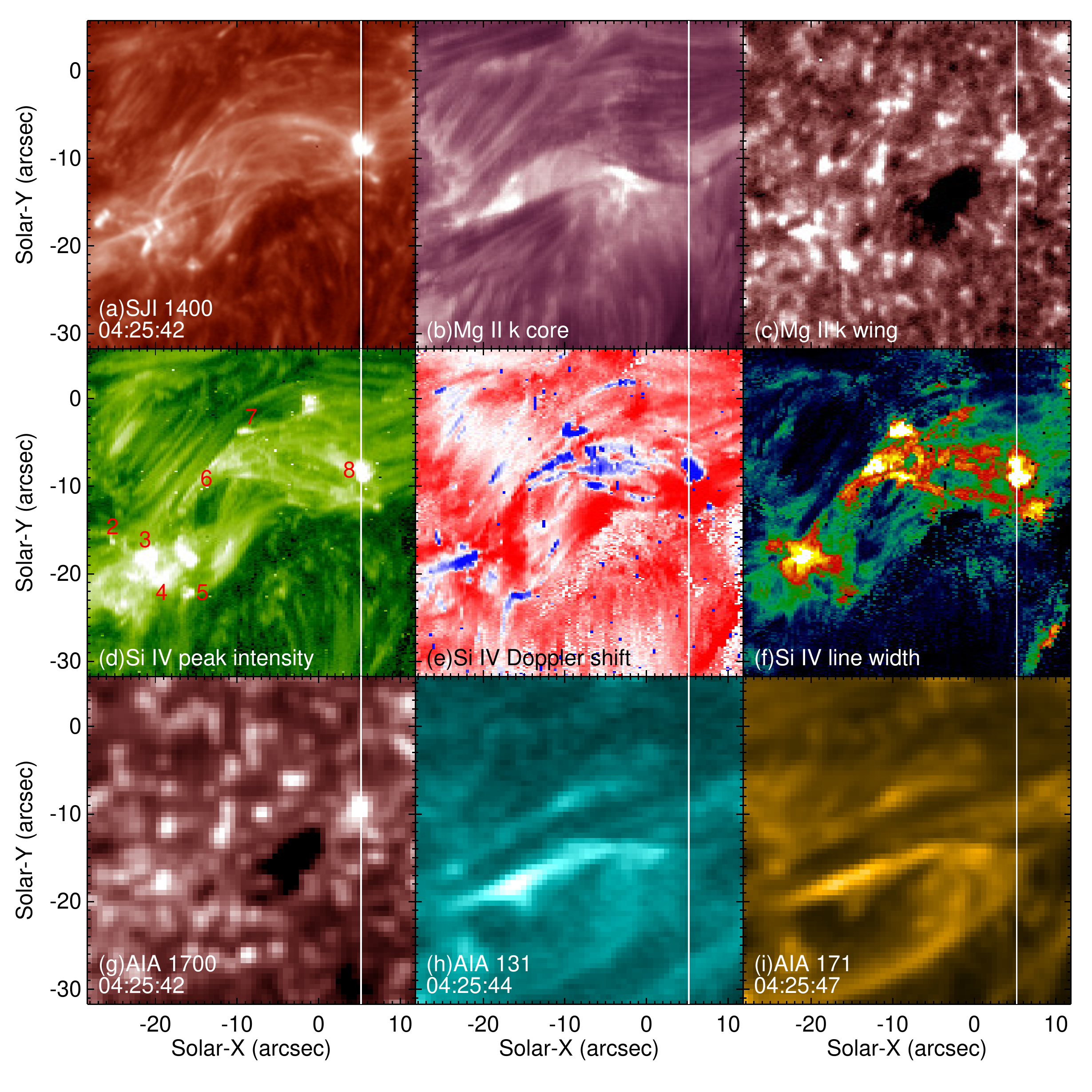}} 
\caption{Similar to Figure ~\ref{f2} but only a smaller region is shown.
} \label{f3}
\end{figure*}

Figure ~\ref{f2} shows an image of IRIS/SJI 1400 {\AA}, and images of AIA 1700, 131, and 171 {\AA} passbands taken around 04:25:42 UT. The Mg~{\sc{ii}} k core and wing images, and Si~{\sc{iv}} 1393.755 {\AA} intensity, Doppler shift and line width images are also shown. The identified UV bursts are marked in Figure ~\ref{f2}(d). Figure ~\ref{f3} presents the same images in a smaller FOV.  We find that the UV bursts have no obvious signatures in the AIA 171 and 131 {\AA} images, which is consistent with previous studies [2,4] and suggests that the UV bursts are likely not heated to coronal temperatures. We can see that many UV bursts appear to be associated with the transition region loops observed in the SJI 1400 {\AA} image. The length of the loops is $\sim$20$^{\prime\prime}$, and these loops may be categorized into cool transition region loops found by Huang et al. [44]. Moreover, some coronal loops can be seen from the AIA 131 and 171 {\AA} images in the same region. It is likely that cool and hot loops are mixed in this region [63,64,65].

\begin{figure*} 
\centering {\includegraphics[width=\textwidth]{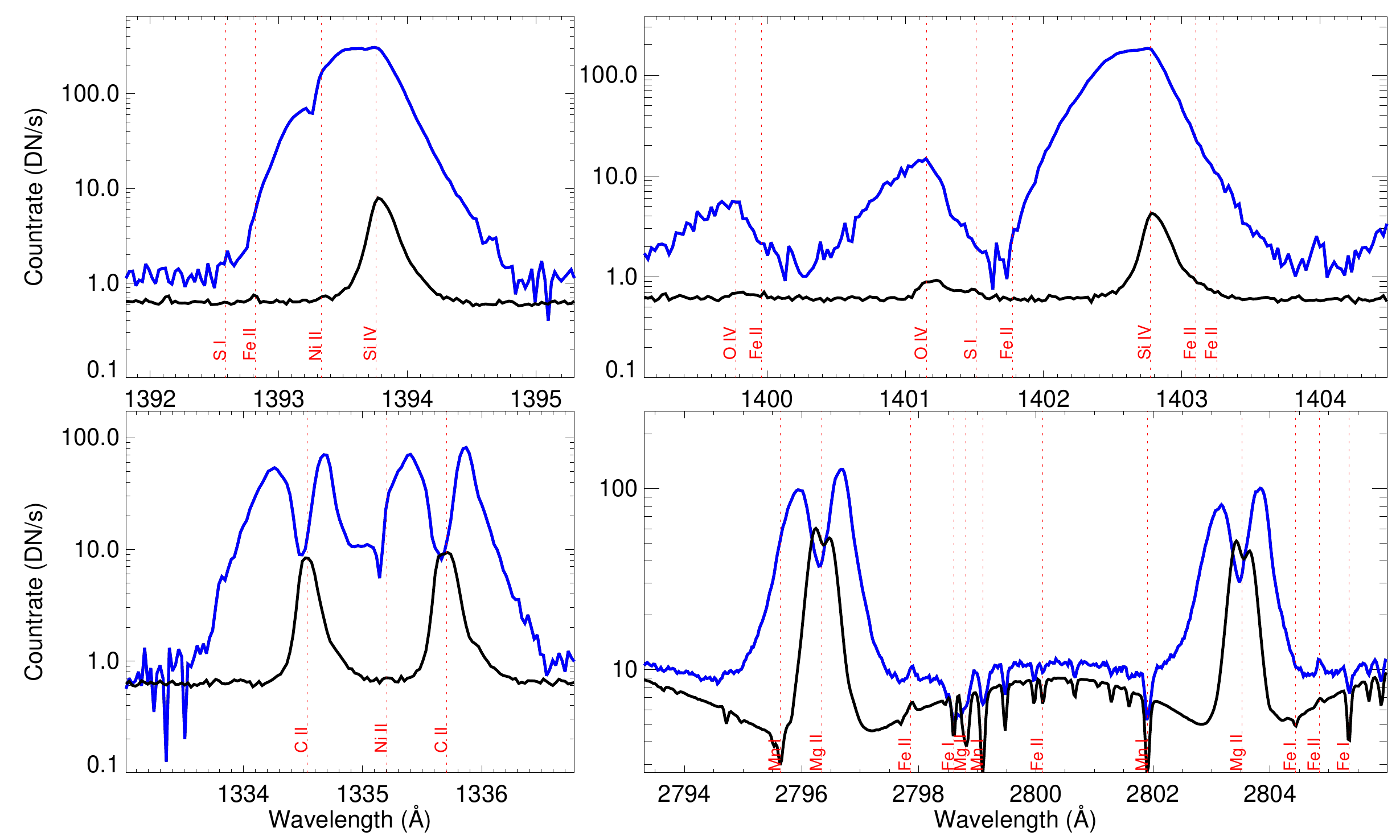}} 
\caption{The blue lines shows the line profiles of burst 3 in four different spectral windows. The black lines present the reference line profiles in a plage region. The vertical dotted lines indicate the rest wavelength of some lines.
} \label{f4}
\end{figure*}

\begin{figure*} 
\centering {\includegraphics[width=\textwidth]{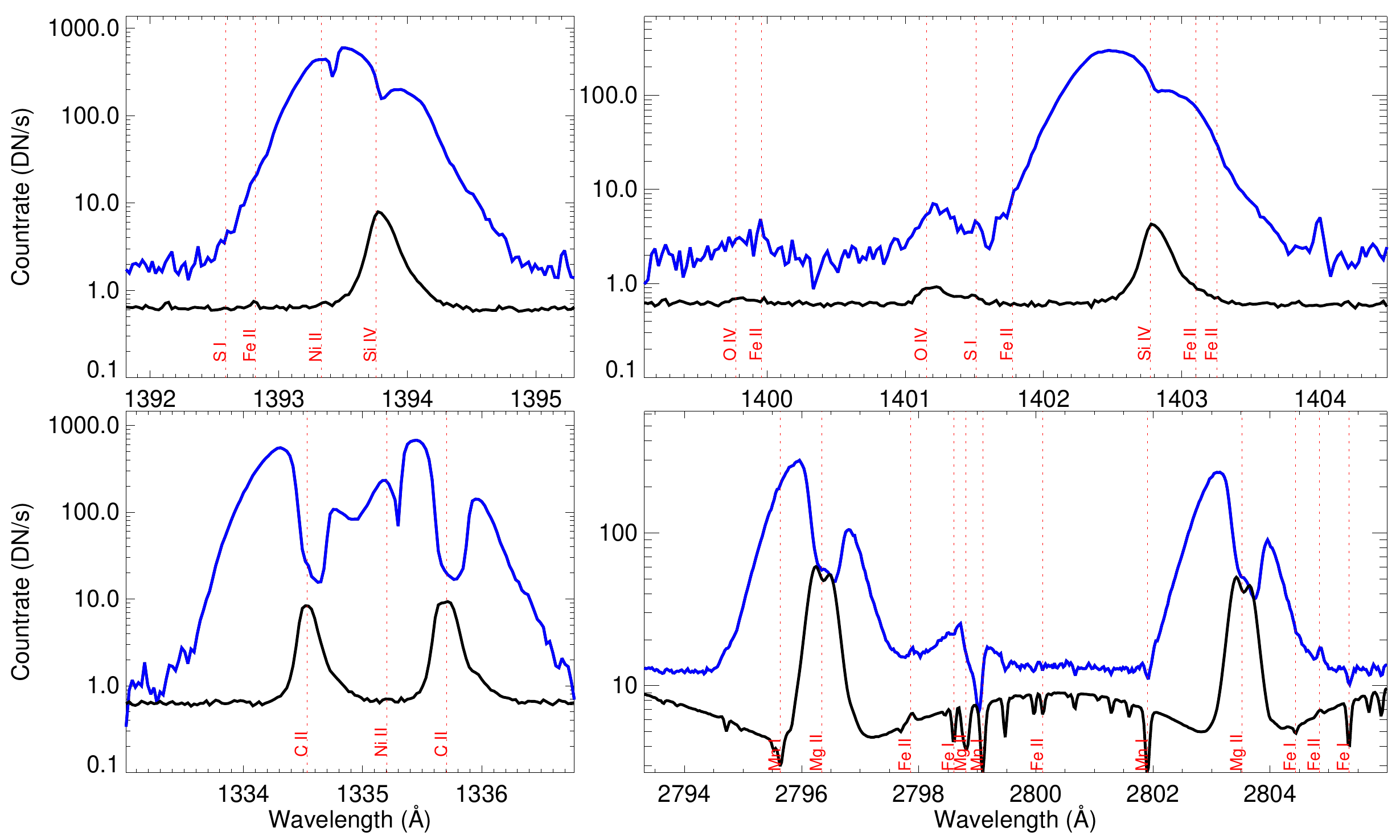}} 
\caption{The same as Figure ~\ref{f4} but for burst 8.
} \label{f5}
\end{figure*}

We present the line profiles of two UV bursts in Figures~\ref{f4} and~\ref{f5}. We can clearly see that the Si~{\sc{iv}} and C~{\sc{ii}} line profiles are greatly enhanced and broadened. The significantly enhanced and broadened wings may correspond to unresolved bidirectional reconnection outflows [2,42]. We also find that UV bursts are mostly associated with large blueshifts of the Si~{\sc{iv}} lines from Figure~\ref{f2}, while the whole region is dominated by redshift. This might indicate that the unresolved upflows dominate over unresolved downflows, which may be related to the density stratification in the atmosphere. It is also possible that the broadened line profiles are caused by the motions of plasmoids with different speeds during reconnection [8,43]. We also find faint dips of the Si~{\sc{iv}} line profiles around the rest wavelengths of the Si~{\sc{iv}} lines in burst 8, as shown in Figure ~\ref{f5}. The dip of the Si~{\sc{iv}} 1393.755 {\AA} line is more prominent than that of the Si~{\sc{iv}} 1402.770 {\AA} line, likely indicating that the dips are caused by self-absorption of the Si~{\sc{iv}} lines instead of the superposition of bi-directional flows. This is because the dips of the two Si IV lines should be similar in case of bi-directional flows. Such a result may indicate that the Si~{\sc{iv}} lines become optically thick during the occurrence of the UV burst. This phenomenon has been reported before [10,45], and dips of the Si~{\sc{iv}} line profiles may be caused by the absorption of overlying transition region loops.

We notice that the Mg~{\sc{ii}} wings and the NUV continuum in burst 8 are greatly enhanced. Burst 8 also shows brightenings in the Mg~{\sc{ii}} k wing image and AIA 1700 {\AA} image, without any obvious signature in the Mg~{\sc{ii}} k core image. In addition, the two O~{\sc{iv}} lines are very weak. This result suggests that burst 8 is likely associated with an EB [4,46]. 

Chromospheric absorption lines usually show a small blueshift in UV bursts, indicating some hotter materials located below the slowly expanding chromosphere [2]. Interestingly, we find that burst 8 reveals a $\sim$20 km~s$^{-1}$ redshift of the Ni~{\sc{ii}} absorption lines (Figure ~\ref{f5}). The redshift of the Ni~{\sc{ii}} absorption lines may correspond to cool materials propagating downwards above the UV burst. We notice that burst 8 is located at one footpoint of the prominent loop system in Figure ~\ref{f3}(a). Bursts 3 and 8 appear to be connected by this loop system. And the Ni~{\sc{ii}} absorption lines at burst 3 are blueshifted by $\sim$15 km~s$^{-1}$(Figure ~\ref{f4}). From the slit-jaw images we do see some materials moving from burst 3 to burst 8 along the loop. It is very likely that these flows are responsible for the redshifted Ni~{\sc{ii}} absorption lines at burst 8. The visibility of the flow in the SJI 1400 {\AA} images and the Ni~{\sc{ii}} lines suggests that the flow contains plasma with temperatures of $\sim$10 kK--100 kK. The speed difference inferred from the Ni~{\sc{ii}} absorption lines in bursts 3 and 8 may be caused by a combination of acceleration by gravity and geometry effect [44].

\section{Magnetic field topologies around the UV bursts}\vspace*{-2mm}

\begin{figure*} 
\centering {\includegraphics[width=\textwidth]{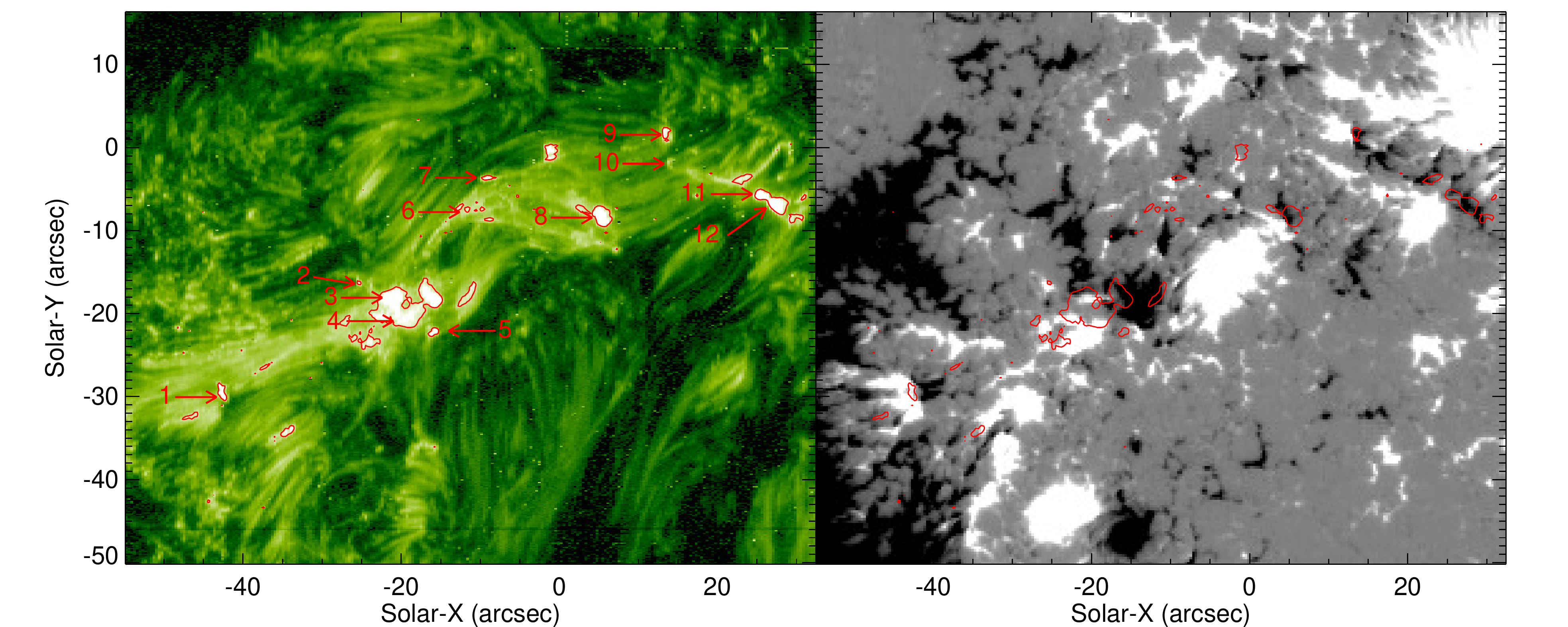}} 
\caption{Left: The image of the Si~{\sc{iv}} 1393.755 {\AA} intensity. Right: \textit{Hinode}/SP line of sight magnetogram. Red contours in both panels represent the brightenings in the Si~{\sc{iv}} intensity image. The intensities are shown in arbitrary unit. The magnetogram saturates at +/$-$ 300 G.
} \label{f6}
\end{figure*}

We have also investigated the magnetic field structures around the UV bursts. Figure ~\ref{f6} shows a comparison between the Si~{\sc{iv}} 1393.755 {\AA} intensity image and the photospheric line-of-sight (LOS) magnetic field taken by \textit{Hinode}/SP. The locations of the UV bursts are marked in the intensity image. Since SP only performed one scan, we can not study the evolution of the photospheric magnetic field. We can see that most of the UV bursts are located at or near regions with strong mixed-polarity magnetic field, which is consistent with previous findings that UV bursts are usually associated with flux cancellations [2,4,7,47]. 

Previous studies have suggested that one particular magnetic field topology, called ``bald patches" [48], are important for UV bursts. Bald patches refer to the dips of serpentine field lines usually formed during flux emergence processes [49,50,51,52,53,54,55,56]. Magnetic reconnection in the lower solar atmosphere can easily occur at bald patches. It is commonly believed that EBs are produced by magnetic reconnection at bald patches [16,51,57,58,59]. Peter et al. [2] also explained the formation of UV bursts using a magnetic field configuration of bald patches. Bald patches have indeed been found at UV bursts [6,7,10]. However, UV bursts are not always associated with bald patches. Chitta et al. [13] found an UV burst associated with the fan-spine magnetic field topology, and that the UV burst evolves together with the corresponding magnetic null point. Also, Tian et al. [10] found that many UV bursts are located at quasi-separatrix layers instead of bald patches.

\begin{figure*} 
\centering {\includegraphics[width=\textwidth]{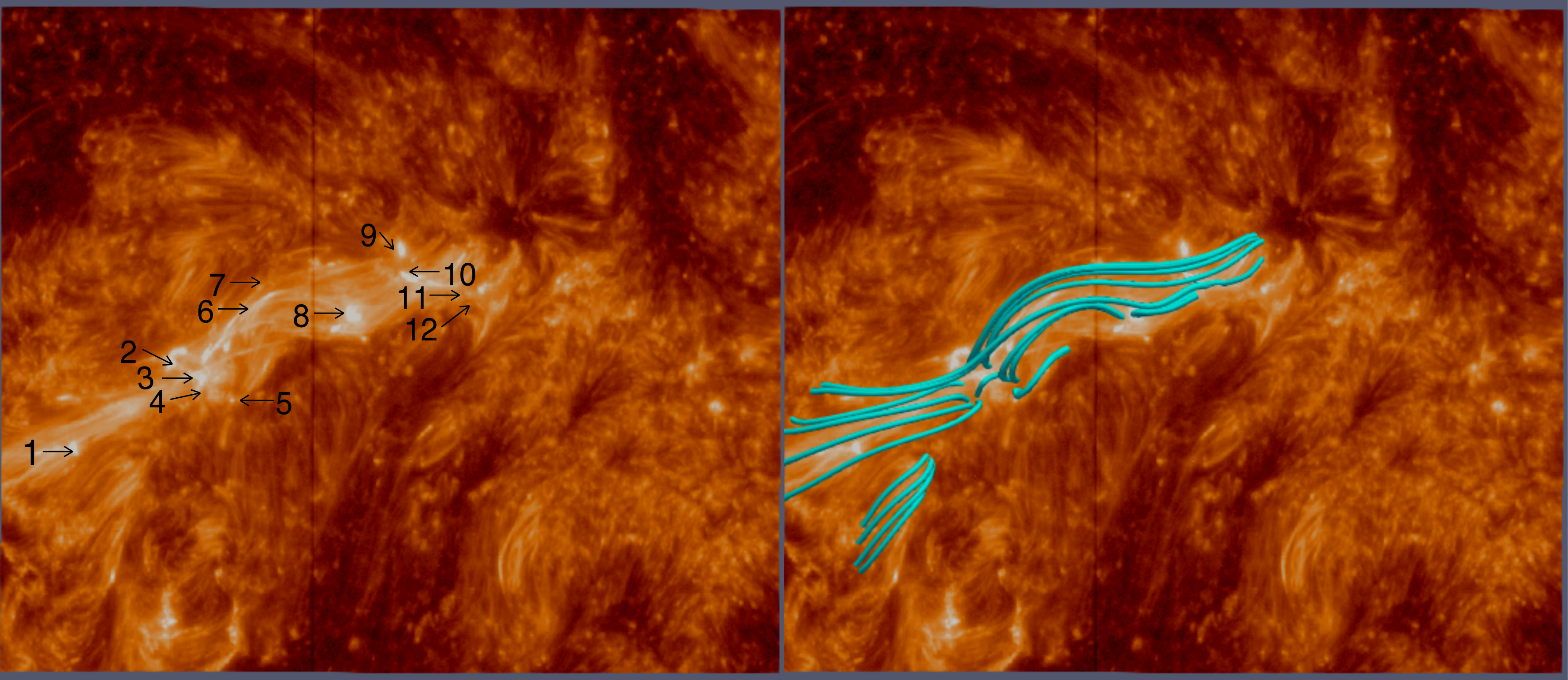}} 
\caption{Comparison between IRIS imaging observation and magnetic field lines from the extrapolation. Left: IRIS/SJI 1400 {\AA} image taken around 04:23:54 UT. Right: Top view of selected magnetic field lines from the extrapolation. The background of the right image is the same as the left image. Identified UV bursts are marked in the left panel.
} \label{f7}
\end{figure*}

In order to investigate the magnetic field topologies around the UV bursts, we extrapolate the photospheric vector magnetic field obtained by SDO/HMI at 04:24 UT to higher heights. Because UV bursts are generally believed to form in the chromosphere or photosphere where the force-free assumption is not valid, force-free magnetic field extrapolation methods may be not applicable here. Thus we build the three-dimensional magnetic field structures using the MHD relaxation method [60,61]. More details of the technique can be found in Tian et al. [10].  Figure ~\ref{f7} presents a comparison between the SJI 1400 {\AA} image taken around 04:23:54 UT and some field lines obtained from the extrapolation. We can see that the extrapolated magnetic field lines match the transition region loops in the SJI 1400 {\AA} image well, indicating that our extrapolation is reasonable. 

\begin{figure*} 
\centering {\includegraphics[width=\textwidth]{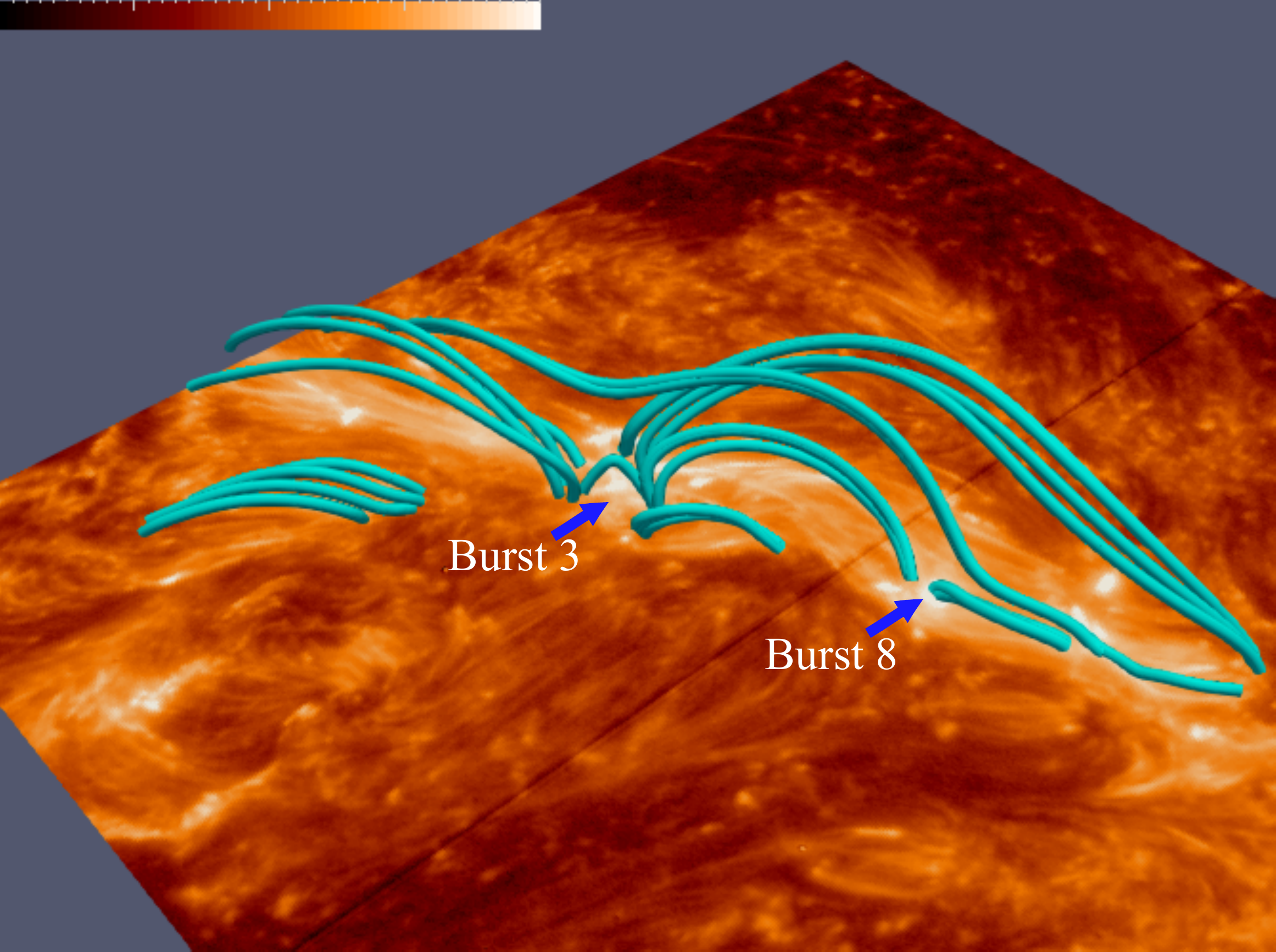}} 
\caption{Similar to the right panel of Figure ~\ref{f7} but with a different viewing angle.
} \label{f8}
\end{figure*}

From Figure ~\ref{f7} we can see that some UV bursts are related to bald patches, while others are not. Figure ~\ref{f8} shows the three-dimensional reconstructed magnetic field from a different viewing angle. We can easily see a bald patch structure around burst 8 and a magnetic null point around burst 3, and these two types of magnetic field topologies are connected by a prominent transition region loop system. Our results confirm that UV bursts could be associated with not only bald patches, but also other magnetic configurations such as magnetic null points.

\section{Coronal signatures in the EIS observation}\vspace*{-2mm}

\begin{figure*} 
\centering {\includegraphics[width=\textwidth]{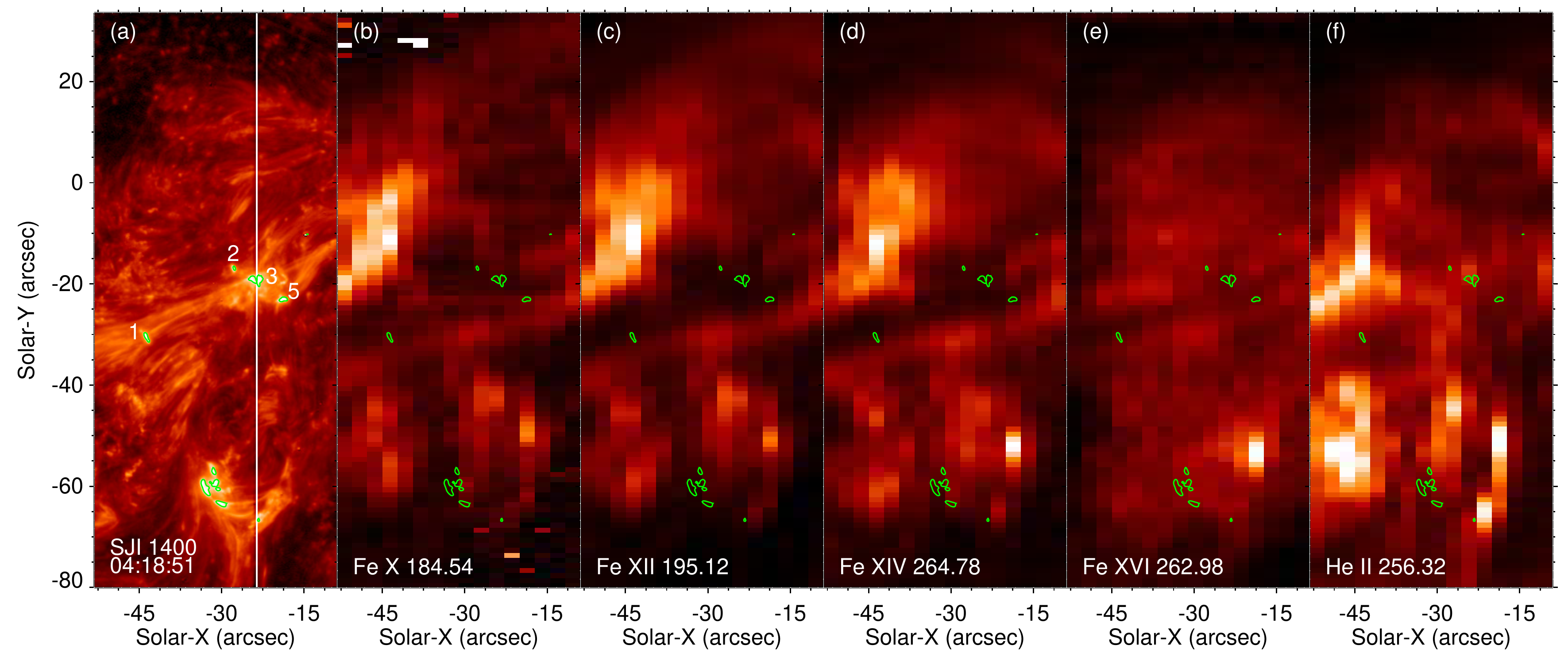}} 
\caption{Coronal signatures of UV bursts in EIS observation.  (a): IRIS/SJI image taken around 04:18:51 UT. The vertical white line indicates the slit location. (b)--(f): Intensity images of different lines obtained from EIS. Green contours outline the brightenings in the SJI 1400 {\AA} image.
} \label{f9}
\end{figure*}

From the \textit{Hinode}/EIS observation, we can examine possible coronal signatures of UV bursts. Slits of EIS and IRIS scanned the same region from 04:10 to 04:23 UT. During this period EIS performed five scans. Some UV bursts have a size of 1$^{\prime\prime}$--3$^{\prime\prime}$ and a lifetime of more than $\sim$4 minutes. The EIS slit successfully scanned several these UV bursts when they are still active. As an example, Figure ~\ref{f9} presents the SJI 1400 {\AA} image taken around 04:18:51 UT, and the peak intensity images of Fe~{\sc{x}} 184.54 {\AA}, Fe~{\sc{xii}} 195.12 {\AA}, Fe~{\sc{xiv}} 264.78 {\AA}, Fe~{\sc{xvi}} 262.98 {\AA}, and He~{\sc{ii}} 256.32 {\AA} taken by EIS at the closest time. The peak intensities are derived from single Gaussian fits. The locations of the UV bursts have been marked in the images.

The intensity images of the coronal lines taken by EIS reveal no obvious brightenings at locations of the UV bursts. There are even no obvious signals of UV bursts in the He~{\sc{ii}} intensity image. We also examine the line profiles at the UV bursts taken by EIS, and do not find any obvious wing enhancement signatures of the coronal lines either (not shown in the paper). As mentioned above, we also could not find any obvious brightenings at the locations of UV bursts in the AIA 131 and 171 {\AA} images (e.g., Figure ~\ref{f2} \& ~\ref{f3}). 

These results indicate that UV bursts are likely not heated to a temperature above 10$^5$ K. Li et al. [38] reported an UV burst with obvious signatures in AIA coronal channels. The UV burst exhibits no absorption features of Ni~{\sc{ii}}. Therefore, the UV burst may occur in the upper chromosphere or transition region, where the plasma can be easily heated to coronal temperatures during magnetic reconnection. As AIA coronal passbands also have some contribution from transition region lines, we can not exclude the possibility that the brigtenings in the AIA coronal images are caused by the transition region emission. Using the pure coronal emission lines observed with EIS, we have demonstrated that there is no coronal signature in the UV bursts identified in our observation. However, it is also possible that the absence of UV bursts in AIA and EIS observations is related to the absorption of emission from possible coronal-temperature plasmas by the He continuum from the overlying cool plasmas.

\section{Summary}

Using IRIS, SDO, and \textit{Hinode} observations of NOAA AR 12280 on 2015 February 7, we have identified 12 UV bursts. The Si~{\sc{iv}} and C~{\sc{ii}} line profiles of these UV bursts are greatly enhanced and broadened, and the Ni~{\sc{ii}} absorption lines are clearly present. We find that the Ni~{\sc{ii}} absorption lines show small blueshifts in most UV bursts. However, in one UV burst the Ni~{\sc{ii}} absorption lines reveal redshifts of $\sim$20 km s$^{-1}$, which appear to be caused by cool materials moving downward above the burst.

We have investigated the three-dimensional magnetic field topologies around the UV bursts using a magnetohydrostatic model. We find both the fan-spine topology and bald patch configuration around the UV bursts.

We have also examined coronal signatures from AIA and EIS observations, and found no obvious brightenings in AIA coronal images and EIS intensity images of coronal lines. The absence of the coronal response suggests that UV bursts are likely not heated to coronal temperatures. But we do not exclude the possibility that possible coronal emission of UV bursts is absorbed by the He continuum in overlying cool plasmas.

\Acknowledgements{\bahao This work is supported by the Strategic Priority Research Program of CAS with grant XDA17040507, NSFC grants 11825301, 11790304 (11790300), 41774183, and 41861134033, and the Strategic Pioneer Program on Space Science of CAS with grants XDA15011000 and XDA15010900. IRIS is a NASA small explorer mission developed and operated by LMSAL with mission operations executed at NASA Ames Research center and major contributions to downlink communications funded by ESA and the Norwegian Space Centre. SDO is the first mission of NASA Living With a Star (LWS) Program. Hinode is a Japanese mission developed and launched by ISAS/JAXA, with NAOJ as a domestic partner and NASA and STFC (UK) as international partners. We thank Dr. Hardi Peter for helpful discussions.}


\normalsize \vskip0.1in\parskip=0mm \baselineskip 18pt
\renewcommand{\baselinestretch}{1.06}\footnotesize\parindent=4mm\bahao

\vskip0.1in \noindent 
\vskip0.1in\parskip=0mm

\REF{1\ }De Pontieu B, Title A M, Lemen J R, et al. The Interface Region Imaging Spectrograph (IRIS). Solar Physics, 2014, 289:2733-2779.
\REF{2\ }Peter H, Tian H, Curdt W, et al. Hot Explosions in the Cool Atmosphere of the Sun. Science, 2014, 346:1255726.
\REF{3\ }Young P R, Tian H, Peter H, et al. Solar ultraviolet bursts. Space Sci Rev, 2018, 214:120.
\REF{4\ }Tian H, Xu Z, He J, et al. Are IRIS bombs connected to Ellerman bombs?. Astrophysical Journal, 2016, 824:96.
\REF{5\ }Fang C, Hao Q, Ding M D, et al. Can the temperature of Ellerman Bombs be more than 10 000 K?. Research in Astronomy and Astrophysics, 2017, 17:31.
\REF{6\ }Toriumi S, Katsukawa Y, Cheung M C M. Various Local Heating Events in the Earliest Phase of Flux Emergence. Astrophysical Journal, 2017, 836:63.
\REF{7\ }Zhao J, Schmieder B, Li H, et al. Observational Evidence of Magnetic Reconnection for Brightenings and Transition Region Arcades in IRIS observations. Astrophysical Journal, 2017, 836:52.
\REF{8\ }Rouppe V D V L, De Pontieu B, Scharmer G B, et al. Intermittent reconnection and plasmoids in UV bursts in the low solar atmosphere. Astrophysical Journal Letters, 2017, 851: L6.
\REF{9\ }Tian H, Yurchyshyn V, Peter H, et al. Frequently Occurring Reconnection Jets from Sunspot Light Bridges. Astrophysical Journal, 2018, 854:92.
\REF{10\ }Tian H, Zhu X, Peter H, et al. Magnetic reconnection at the earliest stage of solar flux emergence. Astrophysical Journal, 2018, 854:174.
\REF{11\ }Scherrer P H, Schou J, Bush R I, et al. The Helioseismic and Magnetic Imager(HMI) Investigation for the Solar Dynamics Observatory(SDO). Solar Physics, 2012, 275:207.
\REF{12\ }Pesnell W D, Thompson B J, Chamberlin P C. The Solar Dynamics Observatory(SDO). Solar Physics, 2012, 275:3.
\REF{13\ }Chitta L P, Peter H, Young P R, et al. Compact solar UV burst triggered in a magnetic field with a fan-spine topology. Astronomy \& Astrophysics, 2017, 605:A49.
\REF{14\ }Ellerman F, Solar hydrogen bombs. Astrophysical Journal, 1917, 46:298
\REF{15\ }Ding M, H\'enoux J C, Fang C. Line profiles in moustaches produced by an impacting energetic particle beam. Astronomy \& Astrophysics, 1998, 332:761.
\REF{16\ }Georgoulis M K, Rust D M, Bernasconi P N, et al. Statistics, Morphology, and Energetics of Ellerman Bombs. Astrophysical Journal, 2002, 575:506.
\REF{17\ }Watanabe H, Kitai R, Okamoto K, et al. Spectropolarimetric Observation of an Emerging Flux Region: Triggering Mechanisms of Ellerman Bombs. Astrophysical Journal, 2008, 684:736.
\REF{18\ }Watanabe H, Vissers G, Kitai R, Rouppe van der Voort L, Rutten R J, Ellerman bombs at high resolution. I. Morphological evidence for photospheric reconnection. Astrophysical Journal, 2011, 736:71.
\REF{19\ }Nelson C J, Shelyag S, Mathioudakis M, et al. Ellerman bombs--evidence for magnetic reconnection in the lower solar atmosphere. Astrophysical Journal, 2013, 779:125.
\REF{20\ }Nelson C J, Scullion E M, Doyle J G, Freij N, Erd\'{e}lyi R, Small-scale structuring of Ellerman bombs at the solar limb. Astrophysical Journal, 2015, 798:19.
\REF{21\ }Vissers G J M, Voort Rouppe van der L, Rutten R J, Ellerman bombs at high resolution. II. Triggering, visibility, and effect on upper atmosphere. Astrophysical Journal, 2013, 774:32
\REF{22\ }Vissers G J M, Voort Rouppe van der L, Rutten R J, et al. Ellerman bombs at high resolution III. Simultaneous observations with IRIS and SST. Astrophysical Journal, 2015, 812:11
\REF{23\ }Yang H, Chae J, Lim E K, et al. Velocities and Temperatures of an Ellerman Bomb and Its Associated Features. Solar Physics, 2013, 288:39.
\REF{24\ }Kim Y H, Yurchyshyn V, Bong S C, et al. Simultaneous observation of a hot explosion by NST and IRIS. Astrophysical Journal, 2015, 810:38.
\REF{25\ }Rutten R J. H$\alpha$ features with hot onsets. I. Ellerman bombs. Astronomy \& Astrophysics, 2016, 590:A124.
\REF{26\ }Fang C , Tang Y H , Xu Z , et al. Spectral Analysis of Ellerman Bombs. Astrophysical Journal, 2006, 643:1325.
\REF{27\ }Bello Gonz\'alez N, Danilovic S, Kneer F. On the structure and dynamics of Ellerman bombs. Detailed study of three events and modelling of H$\alpha$. Astronomy \& Astrophysics, 2013, 557:102.
\REF{28\ }Berlicki A, Heinzel P. Observations and NLTE modeling of Ellerman bombs. Astronomy \& Astrophysics, 2014, 567:A110.
\REF{29\ }Hong J, Ding M D, Li Y, et al. Spectral observations of Ellerman bombs and fitting with a two-cloud model. Astrophysical Journal, 2014, 792:13.
\REF{30\ }Hong J, Carlsson M, Ding M D. RADYN Simulations of Non-thermal and Thermal Models of Ellerman Bombs. The Astrophysical Journal, 2017, 845:144.
\REF{31\ }Hong J, Ding M D, Cao W. Multi-wavelength Spectral Analysis of Ellerman Bombs Observed by FISS and IRIS. Astrophysical Journal, 2017, 838:101.
\REF{32\ }Li Z, Fang C, Guo Y, et al. Diagnostics of Ellerman bombs with high-resolution spectral data. Research in Astronomy and Astrophysics, 2015, 15:1513.
\REF{33\ }Reid A, Mathioudakis M, Kowalski A, et al. Solar Ellerman Bombs in 1-D Radiative Hydrodynamics. Astrophysical Journal Letters, 2017, 835:L37.
\REF{34\ }Ni L, Lin J, Roussev I I, et al. Heating mechanisms in the low solar atmosphere through magnetic reconnection in current sheets. Astrophysical Journal, 2016, 832:195.
\REF{35\ }Ni L, Lukin V S, Murphy N A, Lin J, Magnetic reconnection in strongly magnetized regions of the low solar chromosphere. Astrophysical Journal, 2018, 852:95.
\REF{36\ }Hansteen V H, Archontis V, Pereira T M D, et al. Bombs and Flares at the Surface and Lower Atmosphere of the Sun. Astrophysical Journal, 2017, 839:22.
\REF{37\ }Lemen J R, Title A M, Akin D J, et al. The Atmospheric Imaging Assembly (AIA) on theSolar Dynamics Observatory(SDO). Solar Physics, 2012, 275:17.
\REF{38\ }Li D, Li L, Ning Z. Spectroscopic and imaging observations of small-scale reconnection events. Monthly Notices of the Royal Astronomical Society, 2018, 479:2382.
\REF{39\ }Culhane J L, Harra L K, James A M, et al. The EUV Imaging Spectrometer for Hinode. Solar Physics, 2007, 243:19.
\REF{40\ }Lites B W, Akin D L, Card G, et al. The Hinode Spectro-Polarimeter. Solar Physics, 2013, 283:579.
\REF{41\ }Tsuneta S, Ichimoto K, Katsukawa Y, et al. The Solar Optical Telescope for the Hinode Mission: An Overview. Solar Physics, 2008, 249:167.
\REF{42\ }Innes D E, Inhester B, Axford W I, et al. Bi-directional plasma jets produced by magnetic reconnection on the Sun. Nature, 1997, 386:811.
\REF{43\ }Innes D, Guo L J, Huang Y M, et al. IRIS Si IV Line Profiles: An Indication for the Plasmoid Instability during Small-scale Magnetic Reconnection on the Sun. Astrophysical Journal, 2015, 813: 86.
\REF{44\ }Huang Z, Xia L, Li B, et al. Cool Transition Region Loops Observed by the Interface Region Imaging Spectrograph. Astrophysical Journal, 2015, 810:46.
\REF{45\ }Yan L, Peter H, He J, et al. Self-absorption in the solar transition region. Astrophysical Journal, 2015, 811:48.
\REF{46\ }Grubecka M, Schmieder B, Berlicki A, et al. Height formation of bright points observed by IRIS in Mg II line wings during flux emergence. Astronomy \& Astrophysics, 2016, 593:A32.
\REF{47\ }Nelson C J, Doyle J G, Erd\'elyi R. On The Relationship Between Magnetic Cancellation and UV Burst Formation. Monthly Notices of the Royal Astronomical Society, 2016, 463:2190.
\REF{48\ }Titov V S, Priest E R, D\'emoulin P. Conditions for the appearance of ``bald patches" at the solar surface. Astronomy \& Astrophysics, 1993, 276:564.
\REF{49\ }Fan Y. Nonlinear Growth of the Three‐dimensional Undular Instability of a Horizontal Magnetic Layer and the Formation of Arching Flux Tubes. Astrophysical Journal, 2001, 546:509.
\REF{50\ }Fan Y. The Emergence of a Twisted $\Omega$-Tube into the Solar Atmosphere. Astrophysical Journal Letters, 2001, 554:L111.
\REF{51\ }Pariat E, Aulanier G, Schmieder B, et al. Resistive Emergence of Undulatory Flux Tubes. Astrophysical Journal, 2004, 614:1099.
\REF{52\ }Pariat E, Masson S, Aulanier G. Current Buildup in Emerging Serpentine Flux Tubes. Astrophysical Journal, 2009, 701:1911.
\REF{53\ }Xu Z, Lagg A, Solanki S K. Magnetic structures of an emerging flux region in the solar photosphere and chromosphere. Astronomy \& Astrophysics, 2010, 520:A77.
\REF{54\ }Cheung M C M, Isobe H. Flux Emergence (Theory). Living Reviews in Solar Physics, 2014, 11:3.
\REF{55\ }Schmieder B, Archontis V, Pariat E. Magnetic Flux Emergence Along the Solar Cycle. Space Science Reviews, 2014, 186:227.
\REF{56\ }Danilovic S. Simulating Ellerman bomb-like events. Astronomy \& Astrophysics, 2017, 601:A122.
\REF{57\ }Schmieder B, Rust D M, Georgoulis M K, et al. Emerging Flux and the Heating of Coronal Loops, Astrophysical Journal, 2004, 601:530.
\REF{58\ }Yang H, Chae J, Lim E K, et al. Fine-scale Photospheric Connections of Ellerman Bombs. Astrophysical Journal, 2016, 829:100.
\REF{59\ }Centeno R, Rodr\'iguez, J. Blanco, Del T I J C, et al. A Tale of Two Emergences: Sunrise II Observations of Emergence Sites in a Solar Active Region. Astrophysical Journal Supplement Series, 2017, 229:3.
\REF{60\ }Zhu X, Wang H, Du Z, et al. Forced field extrapolation: testing a magnetohydrodynamic (MHD) relaxation method with a flux-rope emergence model. Astrophysical Journal, 2013, 768:119.
\REF{61\ }Zhu X, Wang H, Du Z, et al. Forced field extrapolation of the magnetic structure of the Halpha fibrils in solar chromosphere. Astrophysical Journal, 2016, 826:51.
\REF{62\ }Ni L, Lukin V S, Murphy N A, Onset of Secondary Instabilities and Plasma Heating during Magnetic Reconnection in Strongly Magnetized Regions of the Low Solar Atmosphere, Astrophysical Journal, 2018, 868:144.
\REF{63\ }Fludra A, Brekke P, Harrison R A, et al. Active Regions Observed in Extreme Ultraviolet Light by the Coronal Diagnostic Spectrometer on SOHO. Solar physics, 1997, 175:487.
\REF{64\ }Spadaro D, Lanzafame A C, Consoli L, et al. Structure and dynamics of an active region loop system observed on the solar disc with SUMER on SOHO.  Astronomy \& Astrophysics, 2000, 359:716.
\REF{65\ }Huang Z. Magnetic Loops above a Small Flux-emerging Region Observed by IRIS, Hinode, and SDO. Astrophysical Journal, 2018, 869:175.
\REF{66\ }Hou Z, Huang Z, Xia L, et al. Narrow-line-width UV Bursts in the Transition Region above Sunspots Observed by IRIS. Astrophysical Journal Letters, 2016, 829:L30.

\end{multicols}

\end{document}